\documentclass[usenatbib]{mn2e} 

\usepackage{psfig}

\def\gsim{\mathrel{\raise0.35ex\hbox{$\scriptstyle >$}\kern-0.6em
\lower0.40ex\hbox{{$\scriptstyle \sim$}}}}
\def\lsim{\mathrel{\raise0.35ex\hbox{$\scriptstyle <$}\kern-0.6em
\lower0.40ex\hbox{{$\scriptstyle \sim$}}}}
\def\gs{\mathrel{\raise0.35ex\hbox{$\scriptstyle >$}\kern-0.6em
\lower0.40ex\hbox{{$\scriptstyle \sim$}}}}
\def\ls{\mathrel{\raise0.35ex\hbox{$\scriptstyle <$}\kern-0.6em
\lower0.40ex\hbox{{$\scriptstyle \sim$}}}}

\def\kms{\,\hbox{km}\,\hbox{s}^{-1}}
\def\Msol{\mathrel{\rm M_{\odot}}}
\def\Wm2{\,\hbox{W}\,\hbox{m}^{-2}}

\begin{document}

\title[A $z\sim0.9$ Supercluster in the UKIDSS/DXS]{The discovery of a massive
  supercluster at z\,=\,0.9 in the UKIDSS DXS}

\author[Swinbank et al.]{
\parbox[h]{\textwidth}{
A.\,M.\ Swinbank,$^{1,*}$
A.\,C.\ Edge,$^{1}$
Ian Smail,$^{1}$
J.\,P.\ Stott,$^{1}$
M.\ Bremer,$^{2}$
Y.\ Sato,$^{3}$\\
C.\ van Breukelen,$^{4}$
M.\ Jarvis,$^{4}$
I.\ Waddington,$^{5}$
L.\ Clewley,$^{4}$\\
J.\ Bergeron,$^{6}$
G.\ Cotter,$^{5}$
S.\ Dye,$^{7}$
J.\ E. Geach,$^{1}$
E.\ Gonzalez-Solares,$^{8}$
P.\ Hirst,$^{9,10}$
R.\,J.\ Ivison,$^{11,12}$
S.\ Rawlings,$^{4}$
C.\ Simpson,$^{13}$
G.\,P.\ Smith,$^{14}$
A.\ Verma,$^{15}$
T.\ Yamada$^{16}$
}
\vspace*{6pt} \\
$^{1}$Institute for Computational Cosmology, Department of Physics, Durham University, South Road, Durham DH1 3LE, UK \\
$^{2}$H.H. Wills Physics Laboratory, University of Bristol, Bristol, BS8 1TL, UK\\
$^{3}$National Astronomical Observatory of Japan, 2-21-1, Osawa, Mitaka, Tokyo 181-8588 Japan\\
$^{4}$Department of Physics, Denys Wilkinson Building, Keble Road, Oxford, OX1 3RH, UK\\
$^{5}$Astronomy Centre, Department of Physics and Astronomy, University of Sussex, Brighton, BN1 9QH, UK\\
$^{6}$Institut d'Astrophysique de Paris, CNRS, 98 bis Bd., Arago, 75014, Paris, France\\
$^{7}$Cardiff University, School of Physics \& Astronomy, Queens Buildings, The Parade, Cardiff, CF24 3AA, UK\\
$^{8}$Institute of Astronomy, University of Cambridge, Madingley Road, Cambridge, CB3 0HA, United Kingdom\\
$^{9}$Joint Astronomy Center, 660 N.\ A.`ohoku Place, Hilo, HI 96720, USA\\
$^{10}$Gemini Observatory, 670 N.\ A`ohoku Place, Hilo, HI 96720, USA\\
$^{11}$UK Astronomy Technology Centre, Royal Observatory, Blackford Hill, Edinburgh, EH9 3HJ, UK \\
$^{12}$Institute for Astronomy, University of Edinburgh, Royal Observatory, Blackford Hill, Edinburgh, EH9 3HJ, UK\\
$^{13}$Astrophysics Research Institute, Liverpool John Moores University, Twelve Quays House, Egereton Wharf, Birkenhead, CH4 1LD, UK\\
$^{14}$School of Physics and Astronomy, University of Birmingham, Birmingham, Edgebaston, B15 2TT, UK\\
$^{15}$Max-Planck-Institut für extraterrestrische Physik, Garching D-85741, Germany\\
$^{16}$Subaru Telescope, 650 N A'ohoku Place, Hilo, HI96720 U.S.A.\\
$^*$Email: a.m.swinbank@durham.ac.uk \\
}

\maketitle

\begin{abstract}
  We analyse the first publicly released deep field of the UKIDSS Deep
  eXtragalactic Survey (DXS) to identify candidate galaxy
  over-densities at $z\sim1$ across $\sim$1 sq. degree in the ELAIS-N1
  field.  Using $I-K$,$J-K$ and $K-3.6\mu$m colours we identify and
  spectroscopically follow-up five candidate structures with
  Gemini/GMOS and confirm they are all true over-densities with between
  five and nineteen members each.  Surprisingly, all five structures
  lie in a narrow redshift range at $z=0.89\pm0.01$, although they are
  spread across 30\,Mpc on the sky.  We also find a more distant
  over-density at $z=1.09$ in one of the spectroscopic survey regions.
  These five over-dense regions lying in a narrow redshift range
  indicate the presence of a supercluster in this field and by
  comparing with mock cluster catalogs from $N$-body simulations we
  discuss the likely properties of this structure.  Overall, we show
  that the properties of this supercluster are similar to the
  well-studied Shapley and Hercules superclusters at lower redshift.
\end{abstract}


%
%
\begin{figure*}
  \centerline{\psfig{file=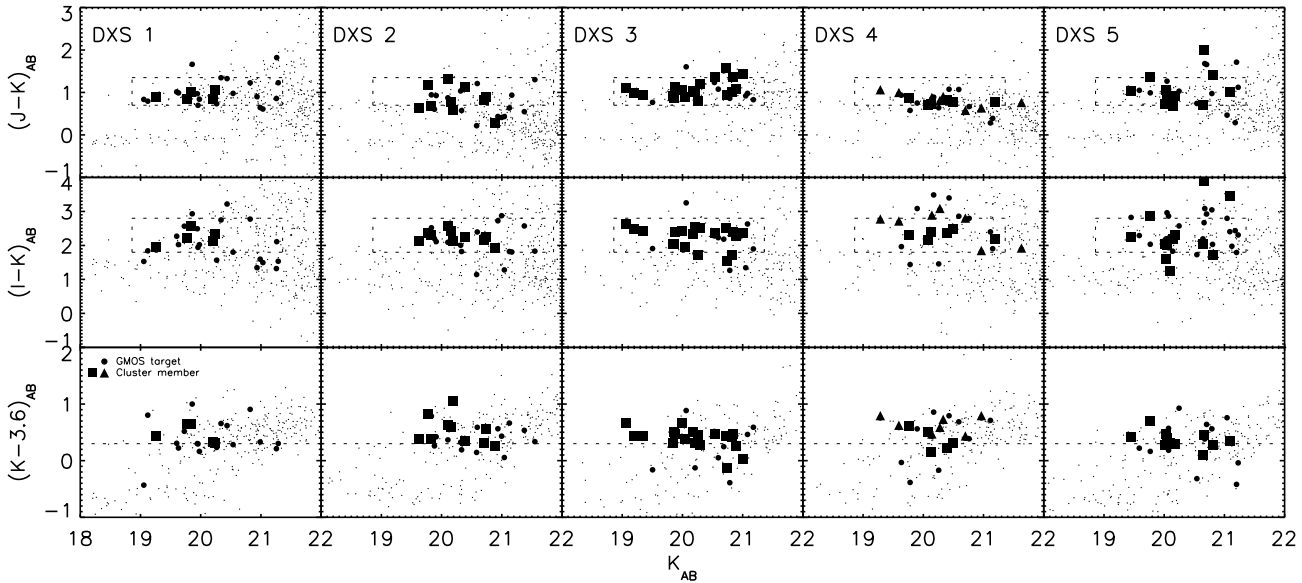,angle=0,width=7in}}
  \caption{The $J-K$, $I-K$ and $K-3.6\mu$m colour-magnitude plots for
    each cluster.  The colour-magnitude diagrams represent a
    1.5\,arcminute region around each cluster.  The dashed lines show
    the limits used to select candidate cluster members via the red
    sequence.  We identify the spectroscopically confirmed $z=$0.89
    cluster members (filled squares), as well as the fore-/background
    galaxies (filled circles).  
    In the DXS4 field we identify separately the members of the higher
    redshift structure in this field (at $z=1.09$).  Note also that
    some spectra were taken for objects outside the colour-magnitude
    limits (marked as dashed boxes) to fully populate the GMOS masks. }
\label{fig:dxs_area}
\end{figure*}

\section{Introduction}

The structure and evolution of clusters of galaxies and their
constituent substructures provides a powerful test of our understanding
of both the growth of large scale structure and dark matter in the
Universe.  In the current hierarchical paradigm of structure formation,
massive galaxy clusters arise from the extreme tail in the distribution
of density fluctuations, so their number density depends critically on
cosmological parameters.  One of the remarkable successes of the
$\Lambda$-CDM paradigm is the match to the number density, mass and
evolution of clusters of galaxies out to $z\sim0.5$--1
\citep{Jenkins01,Evrard02}.  However, due to the very limited number of
clusters known at $z\gsim1$ (where the number density is most sensitive
to the assumed cosmology), the full power of the comparisons to
theoretical simulations has not yet been exploited.

The paucity of clusters known at $z\gsim 1$ stems from the limitations
of current survey methods. For instance, optical colour-selection of
clusters \citep{Couch91,Gladders00}, which relies on isolating the
4000\AA\ break in the spectral energy distributions of passive, red
early-type galaxies (the dominant population in local clusters) becomes
much less effective at $z\gsim0.7$ where this feature falls in or
beyond the $i$-band, in a region of declining sensitivity of
silicon-based detectors.  Recent progress has been made in identifying
clusters using X-ray selection with {\it Chandra} and {\it XMM-Newton}
\citep{Romer01,Rosati02,Mullis05} and these studies have identified
galaxies clusters out to $z\sim 1.5$ \citep{Stanford06,Bremer06}.
However, the X-ray gas in these clusters appears more compact than for
comparable systems at lower redshifts and hence there are concerns that
the accurate comparison of cluster properties with redshift required to
constrain cosmological parameters could be subject to potential
systematic effects related to the thermal history on the intra-cluster
medium.  Thus a complimentary technique for cluster selection is
required.

One solution to this problem is to extend the efficient optical
colour-selection method beyond $z\gsim0.7$ using near-infrared
detectors \citep{Hirst06}.  This approach has been impressively
demonstrated by \citet{Stanford06} who find a $z=1.45$ cluster in the
NOAO-DW survey selected from optical-near-infrared colours.  The
commissioning of the new wide-field WFCAM camera on UKIRT provides the
opportunity to significantly expand deep panoramic surveys in the
near-infrared.  The Deep eXtragalactic Survey (DXS) is a component
within the UK Infrared Deep Sky Survey (UKIDSS) \citep{Warren07} with
the aim of imaging an area of 35 square degrees at high Galactic
latitudes in the $J$- and $K$-band filters to a depth $J_{AB}$=23.2 and
$K_{AB}$=22.7 respectively.  The principal goals of the DXS include
measuring the abundance of galaxy clusters at $z\sim 1$--1.5, measuring
galaxy clustering at $z\sim1$ and measuring the evolution of bias.
This paper presents the first results from a spectroscopic follow-up of
five high redshift galaxy cluster candidates identified in the DXS
Early Data Release (EDR; \citealt{Dye06}).

The structure of this paper is as follows.  In \S2 we describe the data
on which our analysis is based: a combination of optical, near- and
mid-infrared imaging with Subaru, UKIRT and {\it Spitzer} which is used
to select cluster candidates and the follow-up Gemini/GMOS
spectroscopy.  In \S3 we present an analysis of the cluster properties.
We discuss these and give our conclusions in \S4.  Unless otherwise
stated, we assume a cosmology with $\Omega_{m}$=0.27, $\Lambda=$0.73
and $H_{0}$=70$\kms$\,Mpc$^{-1}$.  All magnitudes are given in the AB
system.

\section{Observations and Reduction}

We utilise the UKIDSS-DXS EDR for the ELAIS-N1 region which covers a
contiguous area of $0.86^{\circ}\times0.86^{\circ}$ centred on
$\alpha=16$\,11\,14.400; $\delta=+54$\,38\,31.20 (J2000).  The survey
data products for this region reach 5-$\sigma$ point source limits of
$J_{AB}=22.8$--23.0 and $K_{AB}=22.9$--23.1.  To complement these
observations we exploit deep $I$-band imaging obtained with Suprime-Cam
on Subaru Telescope.  These observations cover the entire ELAIS-N1/DXS
field and are described in Sato et al.\ 2007 (in preparation) and reach
a 5-$\sigma$ point-source limit of $I_{AB}=26.2$.  As part of the {\it
  Spitzer} Wide-area InfraRed Extragalactic (SWIRE) survey
\citep{Lonsdale03}, the ELAIS-N1 region was also imaged in the IRAC
(3.6, 4.5, 5.8 and 8.0$\mu$m) bands as well as at 24$\mu$m with MIPS.
These catalogs are described in \citet{Surace04}.

\subsection{Optical-Infrared matching}

In order to efficiently select cluster candidates, accurate colours are
required for galaxies in the coincident regions of the DXS, Subaru and
SWIRE.  The DXS catalogue was constructed using {\sc SExtractor}
\citep{Bertin96} with a detection threshold of 2$\sigma$ in at least
five pixels.  Objects which lay in the halo or CCD bleed of a bright
star were also removed before the final catalog was constructed.  This
catalogue was then matched to the optical catalogue, with the closest
match within $1''$ being used.  During the first pass cross-correlation
the average offsets between the optical and near-infrared catalogues
was $\Delta \alpha=0.33\pm0.05''$, $\Delta\delta=-0.20\pm0.04''$ (i.e.\
the optical sources were offset to the south-east of the near-infrared
sources, which are tied to FK5 through 2MASS stars, \citealt{Dye06}).
This systematic offset was removed from the optical catalog and the
cross-correlation recalculated resulting in an rms offset of
$\sim0.1''$.  Since accurate colours were required in order to select
cluster candidates, we extracted 2$''$ aperture magnitudes from the
optical and near-infrared catalogs.  In both cases, the magnitude
zero-points were calculated using 2$''$ photometry of (unsaturated)
stars in the field.

The near-infrared and mid-infrared catalogs were cross-correlated in
exactly the same way as above, with a systematic offset between the
mid-infrared and near-infrared sources of $\Delta
\alpha=-0.30\pm0.05''$, $\Delta\delta=0.33\pm0.04''$, which again was
removed before a second pass cross-correlation was performed.

\subsection{Cluster Selection}

As this was a pilot study, we choose to identify candidate
high-redshift galaxy clusters in three ways.  First, we searched for
the sequence of passive red galaxies in high-redshift clusters by
selecting galaxies from photometric catalogue in the $(J-K)$--$K$ and
$(I-K)$--$K$ colour-magnitude space.  We first identified candidates
using slices of $\Delta(J-K)_{AB}=0.4$ stepped between
$(J-K)_{AB}=0$--2.5.  This selection includes a small correction for
the tilt of the colour-magnitude relation for early type galaxies of
$d(J-K)_{AB}/dK_{AB}=-0.025$.  Consecutive slices overlapped by 0.2
magnitudes to ensure that no sequence was omitted.  Each position in
the resulting spatial surface density plot for a colour slice was then
tested for an over-density using consecutively larger apertures from
0.01 to 0.05 degrees (corresponding to approximately 250\,kpc to 1\,Mpc
at $z=1$).  If the over-density in the central aperture was $\geq
3\sigma$ above the background and the density decreased with increasing
aperture radius then a region was marked as a candidate cluster.  A
similar procedure was carried out using the $(I-K)$--$K$
colour-magnitude space to refine the selection of cluster candidates.
Independently, we identified cluster candidates by identifying peaks in
the surface density in $K_{AB}-3.6\mu$m colour space (using an
approximate colour cut of $K_{AB}-3.6\mu{\rm m}>0.3$ which should be
efficient at selecting elliptical at $z\sim$1).  Having defined these
cluster candidates, we checked that each of these met the selection
criterion recently used by \citet{VanBreukelen06} (which is based on a
projected friends of friends algorithm).  Using these three selection
criteria we identified fifteen candidates, of which eight were
identified using all three criteria.  Five of the most promising eight
cluster candidates (which showed the tightest colour-magnitude
sequences and a clear over-density of red objects), were then targeted
for spectroscopic follow-up.  In Fig.~\ref{fig:dxs_area} we show the
colour-magnitude diagrams for a 1.5~arcminute region around each of the
over-density peaks which were spectroscopically targeted (the dashed
boxes in Fig.~\ref{fig:dxs_area} show the colours used to select the
cluster candidates).

%
%
\begin{figure}
  \centerline{\psfig{file=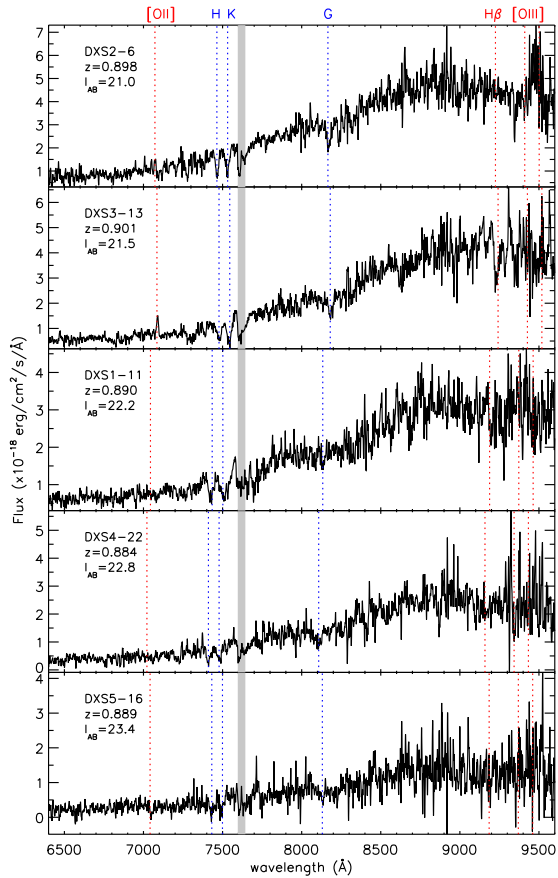,angle=0,width=3in}}
  \caption{Example spectra of the cluster members from the GMOS
    spectroscopy.  The example spectra cover the range of $I$-band
    magnitude of the spectroscopic sample ($I\sim$21.0--23.5).  The
    dashed lines mark the expected position of the emission and
    absorption lines of [O{\sc ii}]3727, Ca\,H\&K, G-band, H$\beta$ and
    [O{\sc iii}]4959,5007.  The grey region shows the position of the
    uncorrected telluric absorption band.}
\label{fig:example_spec}
\end{figure}

%
%
\begin{figure}
  \centerline{\psfig{file=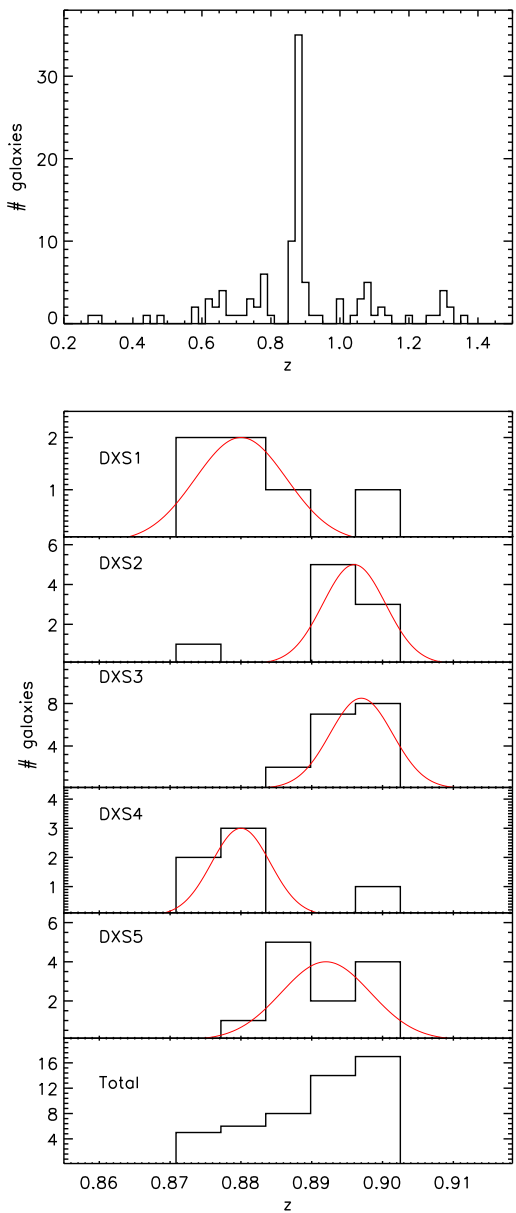,angle=0,width=2.8in}}
  \caption{Top: The total redshift distribution for the five regions
    targeted in our spectroscopic follow-up. This clearly shows the
    strong overdensity in these regions at $z=0.9$.  Middle: The
    redshift histograms for each individual cluster with the mean and
    1-$\sigma$ scatter overplotted (using the method described in
    \S3.2).  The Bottom panel shows the redshift distribution for the
    whole sample on the same velocity scale.  For reference the bin
    size in the lower panels is 1000\,km\,s$^{-1}$ (in the rest frame
    of the cluster).}
\label{fig:histograms}
\end{figure}

\subsection{GMOS Spectroscopy}
\label{sec:GMOSspec}

Spectroscopic follow-up observations of five candidate overdense
regions were taken with the Gemini Multi-Object Spectrograph (GMOS) on
Gemini-North between 2006 May 23 and June 18 U.T.\ in queue mode. As
our target clusters were expected to be at $z\sim 1$ we placed a strong
emphasis on good sky subtraction, to identify weak features in the
presence of strong and structured sky emission.

For this reason we employed the Nod \& Shuffle mode of GMOS.  In Nod \&
Shuffle, the object and background regions are observed alternately
through the same regions of the CCD by nodding the telescope.  In
between each observation the charge is shuffled on the CCD by a number
of rows corresponding to the the centre-to-centre spacing into which
each slit is divided.  Each alternate block is masked off so that it
receives no light from the sky but acts simply as an image store.  The
sequence of object and background exposures can be repeated as often as
desired and at the end of the sequence, the CCD is read, incurring a
read-noise penalty only once (see \citealt{Glazebrook01} for further
details of this general approach).  For each spectrum, the two spectra
block are identified and subtracted to achieve Poisson-limited sky
subtraction.

For our observations we micro-step the targets in the 3.2$''$ long
slits by 1.5$''$ every 30 seconds.  We used the OG515 filter in
conjunction with the R400 grating and a central wavelength of 840\,nm
which results in a wavelength coverage of $\sim580$--1100\,nm.  The
spectral resolution in this configuration is
$\lambda/\Delta\lambda\sim1700$ and the slit width was 1.0$''$.  To
counter the effects of bad pixels and the GMOS chip gaps, the
observations were taken with two wavelength configurations, each
comprising two 2.8-ks exposures at central wavelengths of 840\,nm and
850\,nm respectively.  Each of the five masks was observed for a total
of 3.2 hours in $\lsim0.7''$ seeing and photometric conditions.  In
total 134 galaxies are included on these five masks and we list the
positions and photometric properties ($IJK$ and the IRAC/MIPS bands) of
these in Tables 2\&3.

To reduce the data, we first identified charge traps from a series of
dark exposures taken during the run and used these to mask bad pixels.
We extract the nod and shuffle regions from the data frames and then
mosaiced the three GMOS CCDs.  The frames were then flat-fielded,
rectified, cleaned and wavelength calibrated using a sequence of Python
routines (Kelson, priv. com.).  The final two-dimensional mosaic was
generated by aligning and median combining the reduced two-dimensional
spectra using a median with a 3-$\sigma$ clip to remove any remaining
cosmic rays or defects.  For flux calibration, observations of
BD+28d4211 were taken, however, no tellurics were taken and so we have
not attempted to correct for the $A$-band absorption at 7600\AA\
although this is unimportant for deriving redshifts in any of our
spectra.  While flux-calibration and response correction are not
necessary for redshift determination via cross-correlation, we perform
these steps in order to present the spectra in
Fig.~\ref{fig:example_spec}.

\section{Analysis}

%
\begin{table*}
\begin{center}
{\centerline{\sc Table 1.}}
{\centerline{\sc Properties of the Cluster Candidates}}
\begin{tabular}{lcccccccc}
\hline
\noalign{\smallskip}
Cluster & R.A.\         & Dec.\       & n$_{slits}$ & n$_{cl}$ & $z$        & $\sigma$         & $\sigma$'       \\
        &    \multispan2{(J2000)}     &             &          &            &  (km\,s$^{-1}$)  &  (km\,s$^{-1}$) \\
\hline
DXS1   & 16\,08\,27.0   & +54\,35\,47 & 26          &  5       & 0.8800[19] & $1030\pm270$     & 640$\pm$330     \\
DXS2   & 16\,08\,26.9   & +54\,45\,12 & 25          &  9       & 0.8960[19] &  $700\pm230$     & 440$\pm$170     \\
DXS3   & 16\,09\,05.7   & +54\,57\,23 & 29          & 17       & 0.8970[14] &  $730\pm220$     & 570$\pm$160     \\
DXS4a  & 16\,13\,01.7   & +54\,46\,06 & ...         &  5       & 0.8800[12] &  $660\pm180$     & 470$\pm$230     \\
DXS4b  &                &             & 25          &  8       & 1.0918[14] & $1200\pm340$     &                 \\
DXS5   & 16\,10\,43.6   & +55\,01\,35 & 29          & 12       & 0.8920[20] & $1000\pm290$     & 550$\pm$240     \\
\hline
\hline
\label{table:dxs_clusters}
\end{tabular}
\vspace{-0.5cm}\caption{Names, central positions and properties of the
  spectroscopic sample.  n$_{slits}$ and n$_{cl}$ denote the number of
  spectroscopic slits on the mask and the number of confirmed cluster
  member respectively.  $\sigma$ is the velocity dispersion of the
  spectroscopic sample and $\sigma$' is the velocity dispersion after
  removal of substructure.  The values in the [] denote the errors in
  the last decimal place.}
\end{center}
\end{table*}

\subsection{Redshift determination and velocity dispersions}
\label{sec:redshift}

For redshift determination we first attempt to identify strong emission
or absorption features in the spectra including [O{\sc
  ii}]$\lambda$3726.2,3728.9 emission, the 4000\AA\ break, Ca H\&K
absorption at $\lambda$3933.44,3969.17 or the G-band at
$\lambda$4304.4.  From the sample of 134 galaxies with spectroscopic
observations, 111 yielded secure redshifts, with only 23
unidentifiable, giving a 85\% success rate.  As expected for
absorption-line spectroscopy, the non-detections can in large part be
attributed to optical faintness: the median $I_{AB}$ magnitude of the
111 galaxies with secure redshifts is $22.15\pm0.2$, whereas for the
galaxies without redshifts the median magnitude was
$I_{AB}=22.65\pm0.4$.  The measured redshifts for all sources are
listed in Tables 2\&3.

Having identified an approximate redshift for a galaxy we compute a
robust velocity by cross-correlating each the spectrum with an
elliptical galaxy template spectrum.  For the template we use solar
metallicity, 1\,Gyr burst models with ages of 3,5 or 7\,Gyr from
\citet{Bruzual03}.  The errors on the redshifts are determined from the
shape of the cross-correlation peak and the noise associated with the
spectrum and are typically in the range 30--150\,km\,s$^{-1}$. We
present typical example spectra from each of the masks in
Fig.~\ref{fig:example_spec} and report the errors on the redshifts of
the individual candidate cluster members in Table~2.

Figure~\ref{fig:histograms} shows the redshift distribution for all
galaxies in our sample.  We see that the vast majority of the
spectroscopic sources lie in a narrow redshift range at $z\sim 0.9$.
This indicates that most of the galaxies we have selected lie within
the overdense regions we targeted and that these structures themselves
appear to form a coherent structure across the whole survey region.

%
%
\begin{figure*}
  \centerline{\psfig{file=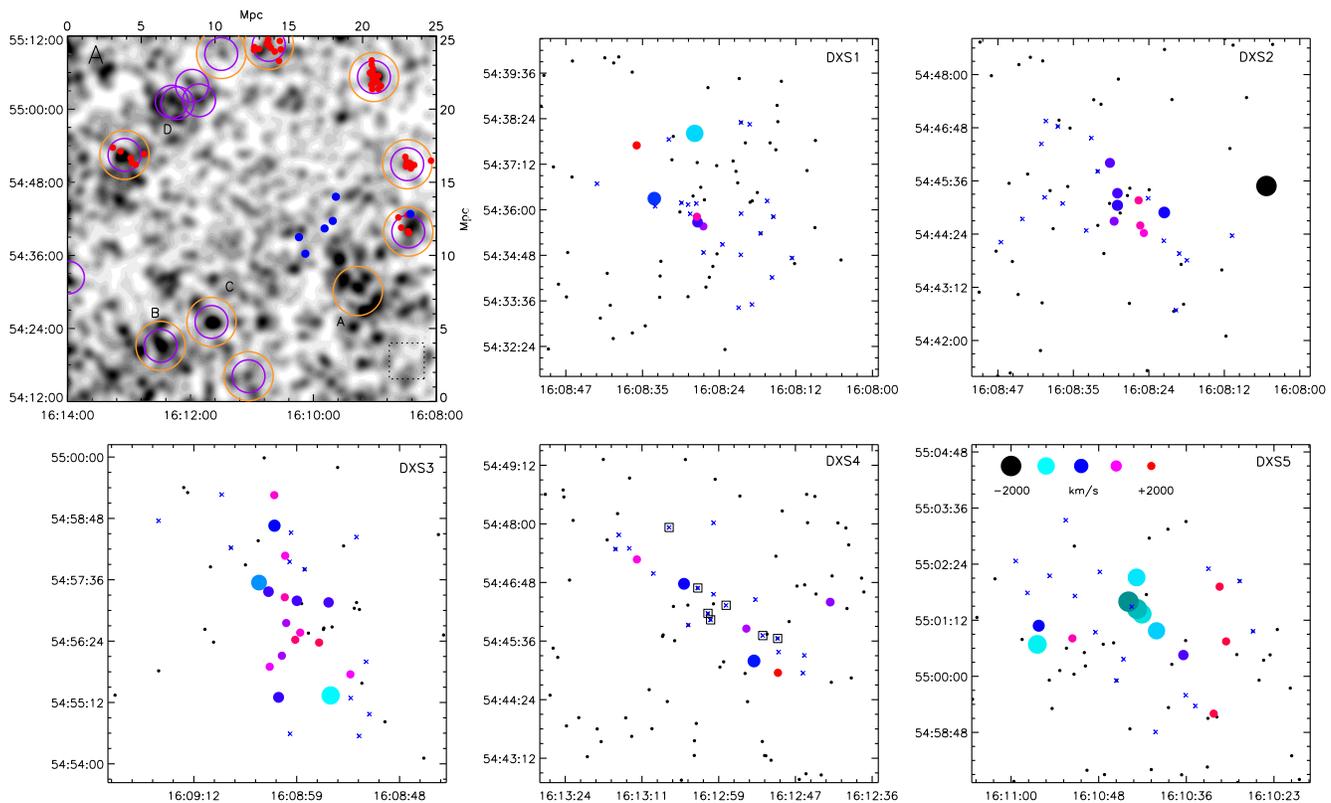,angle=90,width=7in}}
  \caption{{\it Panel A:} Colour-selected surface density map of the
    ELAIS-N1 region based on the $I-K$, $J-K$ and $K$-3.6$\mu$m colours
    used to select galaxy clusters (the cluster candidate selection is
    described in \S2).  The image is smoothed with a Gaussian kernel
    with a FWHM of $60''$ (420\,kpc at $z=0.9$).  The large open
    circles represent the cluster candidate colour selection using
    $J-K$ (large circles) and $K$-3.6$\mu$m (slightly smaller) colours.
    The dashed square box in the bottom right hand corner shows the
    size of the GMOS field of view.  For the five regions targeted in
    our GMOS observations we mark the individual galaxies which are
    known to be members and in addition we plot those galaxies from the
    literature which lie between $z=0.870$ and $z=0.915$.  The cluster
    candidates marked A-D have colour-magnitude sequences consistent
    with the z=0.89 supercluster.  It is clear that the supercluster
    potentially spans the whole field (the most prominent region is at
    16:10:09, 54:25:00) and beyond.  {\it Panels DXS1-5:} The spatial
    distribution of the galaxies within in each of the five over-dense
    regions selected for spectroscopic study.  We plot the positions of
    all of the galaxies which meet our colour-selection (see \S2.2) and
    we identify those which are spectroscopically confirmed as cluster
    members or non-members.  For the members, the sizes and colours of
    the symbols denote the rest-frame velocity offset with respect to
    the cluster redshift given in Table~1 (on a velocity scale from
    $-2000$ to $+2000$\,km\,s$^{-1}$).  In addition, in the DXS4 field
    we identify the members of the background $z=1.09$ structure.  }
\label{fig:area_vel}
\end{figure*}

To calculate the membership of the structure in each field, we have
followed the iterative method used by \citet{Lubin02}.  Initially we
estimated the central redshift for the overdensity, and select all
other galaxies with $\Delta z < \pm0.06$ in redshift space.  We then
calculated the bi-weight mean and scale of the velocity distribution
\citep{Beers90} which correspond to the central velocity location,
$v_{c}$, and dispersion, $\sigma_{v}$ of the cluster. We used this to
calculate the relative radial velocities in the restframe: $\Delta v =
c(z-z_{c})/(1+z_{c})$. The original distribution was revised, and any
galaxy that lies $>3\sigma_{v}$ away from $v_c$, or has $| \Delta {v} |
> 3500$\,km\,s$^{-1}$ was rejected from the sample and the statistics
were re-calculated.  The final solution is achieved when no more
galaxies are removed by the iterative rejection. The results are
presented in Table~1, with 1-$\sigma$ errors on the cluster redshift
and dispersion corresponding from $10^{3}$ bootstrap re-samples.  We
plot the redshift histograms for the structures in each field in
Figure~\ref{fig:histograms} along with a Gaussian curve showing the
measured mean redshift and velocity dispersion.

The most striking result from these histograms is the discovery that
all five structures lie within 3000\,km\,s$^{-1}$ of each other even
though they are spread across nearly a degree on the sky (approximately
30\,Mpc in projection).  This strongly suggests that this field
intercepts a ``supercluster'' like structure at $z=0.9$ -- we discuss
the posterior likelihood of this in \S\ref{sec:discussion} and next
discuss the the properties of the individual structures.

\subsection{Searching for Substructures}

\begin{figure}
  \centerline{\psfig{file=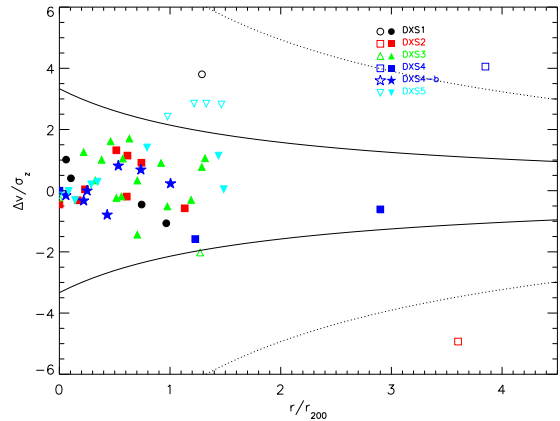,angle=90,width=3in}}
  \caption{The cluster membership technique of Carlberg, Ellingson \&
    Yee (1997) applied to our cluster sample.  The solid curve denotes
    the 3$\sigma$ contour of the mass model; the dashed curve is the
    6$\sigma$ contour.  The filled symbols denote galaxies which lie
    within the 3$\sigma$ contours whilst the open symbols denote
    galaxies which lie outside the 3$\sigma$ contours. This test shows
    that significant substructure is evident in DXS5, with outliers
    also evident in DXS1, 2 \& 3.  Restricting our analysis to the
    galaxies within the central 3$\sigma$ limits we derive velocity
    dispersions for the whole sample of 540$\pm$100$\kms$.}
\label{fig:DvDr}
\end{figure}

We list the centres of the structures identified from our dynamical
analysis in Table~1, along with the number of members, the mean
redshift and the estimate of the velocity dispersion for each
structure.  Since the central positions and redshifts of the clusters
are not well constrained we define the central velocity as the median
redshift in the cluster and determine the centre of the cluster from
the peak in the cluster surface density plot (Fig.~\ref{fig:area_vel}).
The uncertainties on the velocity dispersions are derived from
bootstrap resampling the observed sample of members.  Measuring the
velocity dispersions from clusters with $\sim$10 members is
particularly difficult, and in the rest-frame, the cluster velocity
dispersions are unusually high ($\sim$1000$\kms$).  We derive more
secure velocity dispersions by first investigating how relaxed each of
the structures are, and construct position--velocity diagrams (similar
to Dressler--Shectman plots; \citealt{Dressler88}).  In
Figure~\ref{fig:area_vel} we mark the positions of all of the galaxies
for which a radial velocity measurement was obtained (we note that the
flat distribution of galaxies in this plot reflect the spatial sampling
by GMOS).  Together with Fig.~\ref{fig:histograms}, this shows that the
only structure with a discernible non-gaussian velocity distribution is
DXS5, where four galaxies form a higher velocity substructure.  As
noted above, unfortunately, the small numbers of members in each
structure compromise the conclusions we can draw from this analysis.
However, we can attempt to derive average velocity dispersions from the
whole sample.  We de-redshift and stack the five clusters (according to
their central redshifts) and measure a velocity dispersion of
900$\pm$200$\kms$ which may remain artificially high due to
substructure.  In order to better define the cluster membership via a
simple method we use both the velocity and spatial information.  This
technique was first used in the CNOC surveys \citep{Carlberg96} 
and is described in detail in \citet{Carlberg97c} and briefly described
here.  Firstly, the mean redshift of the cluster is normalised to the
observed velocity dispersion ($\sigma_{z}$).  This is plotted against
the projected radius away from the center of the cluster in units of
$r_{200}$ (Fig.~\ref{fig:DvDr}).  The mass model of \citet{Carlberg97c}
can then be used to mark the 3$\sigma$ and 6$\sigma$ limits which are
used to differentiate between cluster members and near-field galaxies
(or galaxies which reside in filaments/structures surrounding the
clusters).  In this analysis $r_{200}$ is calculated under the
assumption that a cluster is a single isothermal sphere and is defined
to be the clustocentric radii at which the mean density interior is
200$\times$ the critical density at the redshift of the cluster.  We
calculate $r_{200}$ as $r_{200}$ = $\sqrt{3}\sigma_{z}$/10$H(z)$. where
$H(z)$ is defined by $H(z)=H_{0}^{2}(1+z)^{2}(1+\Omega_{0}z)$ (see
\citet{Carlberg97c} for a detailed discussion).  Restricting our
analysis to the galaxies which lie within the 3$\sigma$ limits we
recalculate the velocity dispersion for the clusters as an ensemble and
derive a velocity dispersion of 540$\pm$100$\kms$.

Given the limited number of cluster members, it is not practical to
establish accurate limits on the fraction of velocity substructures.
However, using the analogy with the Shapley supercluster which has
several multi-component velocity clusters (e.g.\ A\,1736 and A\,3528),
we can state that the observed fraction of 20\% in the five DXS
clusters is consistent with other superclusters at low redshift
(although this clearly suffers from small number statistics).

\subsection{Spectral Classification}

To investigate the spectral properties of the cluster galaxies, we
spectroscopically classify the galaxies in our sample according to the
classification of \cite{Dressler99}.  We find an average spectral mix
of: k, $29\pm7$\% ($17\pm4$); k+a: $24\pm8$\% ($14\pm5$); e(a):
$17\pm3$\% ($10\pm2$); e(c): $28\pm7$\% ($16\pm4$); e(b): $2\pm2$\%
($1\pm1$).  The structure in DXS3 (and to a lesser extent DXS5) show
excess numbers of active galaxies (e(c) and e(a)) compared to the other
fields, but these are only marginally significant.  These spectral
mixes are similar to previous spectroscopic studies of similarly
high-redshift galaxy cluster members (e.g. \citealt{Jorgensen05}) as
well as that seen in local ($z\sim0.1$) galaxy clusters (e.g.
\citealt{Pimbblet06}) which have found that the mix of k and k+a
galaxies make up 50-70\% of the population whilst star-forming galaxies
contribute $\sim$20\% with the remaining having properties consistent
with e(a) galaxies.  The most significant difference is that the DXS
clusters have up to 20\% of galaxies with e(a) signatures, which is
slightly higher than local clusters or other high-redshift clusters.

We note that there is also a large dispersion in the fraction of [O{\sc
  ii}] detections in the cluster members of each of the six clusters.
In total there are 56 cluster members, of which 27 have significant
[O{\sc ii}] emission ($48\pm9$\% with an equivalent width above 3\AA)
which is comparable to that found in similar $z>0.6$ clusters
\citep{Finn05,Poggianti06}.  However, this global fraction hides a wide
range in the fraction from cluster to cluster (between 20 and 75\% for
DXS4 to DXS3 respectively).  While the median $I-K$ colour of members
with or without [O{\sc ii}] are indistinguishable (both
$I-K=2.41\pm0.28$), this may be due to the fact that the $I$-band
data cover rest-frame emission redward of the 4000\AA\ break.

In terms of the 24$\mu$m detections, we note that of the six galaxies
which have 24$\mu$m counterparts in the clusters, two have spectral
properties consistent with passive galaxies (k+a), three are strongly
star-forming (with strong [O{\sc ii}], H$\beta$ and [O{\sc iii}]
emission lines) and one galaxy (DXS4-11) shows high excitation lines
(such as [Ne{\sc v}]3346,3426 and [Ne{\sc iii}]3343,3868.7) which
unambiguously identifies this source as a highly-obscured AGN.

\section{Discussion}
\label{sec:discussion}

The most striking result from our survey is the discovery of five
clusters at $z$=0.89 across 30\,Mpc in projection.  The velocity
dispersions and physical sizes of each of these individual clusters
bear a number of similarities to (well studied) local superclusters,
and therefore lead us to interpret the results in the context of a
``super-cluster'' at $z$=0.89.

\subsection{How much of the supercluster have we identified?}

To investigate how much of the supercluster remains unidentified (since
only the first five cluster candidates we spectroscopically targetted),
we colour cut the ELAIS-N1 catalog and construct a surface density plot
to look for other potential supercluster members.  Using the colour
cuts $K=18.8$--21.3; $(J-K)=0.7$--1.35; $(I-K)=1.95$--2.95 and
$K-3.6\mu$m$>0.3$; (see Fig.~\ref{fig:dxs_area}) we construct a
colour-selected density map of the ELAIS-N1 region and present the
results in Figure~\ref{fig:area_vel}.  We also overlay the
spectroscopically identified cluster members from DXS1-5 and objects
from previous studies that have spectroscopically identified
$z\sim$0.90 galaxies in this field
\citep{Scott00,Chapman02b,Manners03}.

This surface density map identifies all the candidate clusters we
selected.  Of the seven cluster candidates we did not observe, four
(marked A-D in Fig.~\ref{fig:area_vel}) have colour sequences
consistent with a cluster at $z=0.90$ (the other candidates have
colour-magnitude sequences which are likely lower redshift).  We also
note that since all five supercluster members are close to the edge of
the WFCAM field, we may have only partially sampled the full
supercluster.  Although we currently do not have the near-infrared
imaging outside the $0.8\times0.8$\,degree field to efficiently select
other supercluster candidates we estimate that we may have missed up to
50\% of the full structure.  Therefore, there are potentially between 7
and 18 rich clusters in the supercluster on a scale of 50--60\,Mpc.
This is consistent with local superclusters such as Shapley and
Hercules (which have seven and nine Abell class two or above clusters
in a redshift range equivalent to 4000$\kms$ across $\sim$60\,Mpc; e.g.
\citealt{Barmby98}).  Thus, the observations presented here highlight
the need to study fields on scales of several degrees to best
characterise such large structures even at $z\sim1$.

\subsection{How rare are superclusters?}

The discovery of a massive supercluster in the first DXS survey field
of nearly fifty is surprising given their low space density below
$z$=0.1.  Cluster surveys at higher redshift have also identified
superclusters similar to the structure presented here (e.g.\
Cl\,1604+4321 $z=0.90$, \citealt{Gal04}) and quasar surveys have
revealed massive overdensities at still higher redshift
\citep{Graham96}.  Therefore, this system is not unique but it is
important to estimate how likely it is that we should have identified
one in the first UKIDSS-DXS field.

Using the statistics from \citet{Tully86,Tully88} for the local space
density of superclusters, we estimate that there are five superclusters
over the high galactic latitude sky $within~z=0.1$.  This corresponds
to one supercluster per 0.04\,Gpc$^3$.  The total volume sampled in the
complete DXS survey between redshifts 0.7 to 1.4 (the farthest we can
efficiently select galaxy clusters from the DXS and reliably recover
redshifts for from optical spectroscopy) will be 0.27\,Gpc$^3$
(comoving).  Scaling from the local space density of superclusters, we
expect a total of seven superclusters in the complete DXS survey. To
find one such system in the first field from the DXS is fortunate
($\sim$15\% probability) but not so unlikely for us to question the
validity of our interpretation of this system as a rich supercluster.

To estimate the potential masses of this structure and the clusters
within it, we compare the observed space density of clusters in the
ELAIS-N1 region to the predictions of the expected space density of
dark matter halos from the $N$-body simulations of \citet{Reed05}.
Within the survey volume covered by this DXS field, the halo mass
functions from \citet{Reed05} suggest that at $z$=0.9 there should be
$\sim$60, 25, 2 and 0.05 halos of mass log$(M/\Msol$)=13,13.5,14.0 and
15.0 respectively within our survey volume of
3$\times$10$^{6}$Mpc$^{3}$.  Thus, the number density of massive halos
found in the region we have surveyed is consistent with halos of mass
$\sim$10$^{13.5-14.0}\Msol$.

However, these estimates disregard clustering of clusters and so we
also exploit the Hubble Volume cluster catalog \citep{Evrard02} which
uses giga-particle $N$-body simulations to study galaxy cluster
populations in CDM simulations out to $z=$1.4.  We exploit the NO sky
survey catalog of the $\Lambda$CDM cluster simulation which covers a
redshift from $z=$0 to $z=$1.5 in a solid angle of $\pi$/2 steradians.
We randomly sample this catalog in volumes comparable to the DXS
ELAIS-N1 survey, and find that the probability of finding five clusters
with velocity dispersions $\gsim$450$\kms$ between $z=$0.7 and $z=$1.4
is $\sim$75\%.  However, the chance of all five clusters lying within a
2000$\kms$ slice is only $\sim$20\%.  When this criterion is met, we
note that median mass of each cluster is $\sim10^{13.5-14.0}\Msol$
(crudely suggesting a total mass for five clusters $>10^{14.7}\Msol$).

\section{Conclusions}

We present the results of the first spectroscopic follow-up of
candidate high-redshift clusters selected from the UKIDSS DXS.  This
pilot programme was designed to test the feasibility of identifying
high-redshift ($z$=0.8--1.4) galaxy clusters in the first DXS survey
field through an extension of the red-sequence method which efficiently
selects galaxy clusters at $z\lsim$0.7 \citep{Gladders00}. The main
results are summarised as follows:

\begin{enumerate}

\item Using ($J-K$), ($I-K$) and ($K$-3.6$\mu$m) colours we extend the
  efficient red-sequence cluster detection method developed by
  \citet{Gladders00} and identify fifteen cluster candidates in the 0.8
  square degree DXS ELAIS-N1 field.  Five cluster candidates were
  targetted with GMOS spectroscopy, all of which yielded significant
  overdensities between $z$=0.88 and $z$=1.1 with between five and
  nineteen members.  The 100\% success rate of this cluster search
  confirms that the colour selection is efficient at selecting the
  highest redshift galaxy clusters.

\item The most striking result from our observations is that five of
  the six galaxy clusters lie within 3000$\kms$ of each other across
  30\,Mpc in projection.  This overdensity is most naturally explained
  by the presence of a supercluster at $z$=0.9 (at least part of) which
  our observations intersect.

\item The clusters have velocity dispersions of between 600 and
  1200$\kms$, although by removing substructures, we derive velocity
  dispersions of 540$\pm$100$\kms$ (consistent with individual halos of
  mass $\sim10^{13.5-14.0}\Msol$).

\item We find that the mix of $k$ and $k+a$ galaxies make up 50-70\% of
  the population, whilst star-forming galaxies comprise 20\% with the
  remaining galaxies having properties consistent with e+a signatures.
  The spectroscopic mix of galaxies is similar to previous studies of
  both low- ($z\sim$0.1) and high- ($z\sim$0.7) redshift clusters.  We
  also derive redshifts for six cluster 24$\mu$m sources, two of which
  are passive ($k+a$), three strongly star-forming (strong [O{\sc ii}],
  [O{\sc iii}] and H$\beta$) and one with high excitation lines
  indicating an AGN.

\item By comparing the number of clusters in our survey volume with the
  number density of massive halos from N-body simulations we suggest
  that each of these clusters will have masses of order
  $10^{13.5-14.0}\Msol$.  Moreover, we also compare the cluster abundance
  with predictions from giga-particle $N$-body simulations to estimate
  the probability of finding such a structure as $\sim$25\% in the
  current cosmological paradigm.

\end{enumerate}

Whilst simulations and mock cluster catalogs provide extremely useful
constraints on the likelihood of finding such a structure and crude
estimates of the mass, the ultimate goal of the complete DXS survey
area (35 square degrees) is to measure cluster abundances between
$z\sim$0.8-1.4.  In order to constrain cluster abundance, a combination
of follow-up spectroscopy and sophisticated mock catalogs will be
required in order to accurately constrain the masses of the clusters
from galaxy velocity dispersions (e.g\,\citealt{Eke06}).  If reliable
halo masses can be derived, then for a fixed set of cosmological parameters
($\Omega_{m}$, $\Omega_{\Lambda}$), the resulting cluster abundance
will reflect on $\sigma_8$ (the rms mass fluctuation amplitude in
spheres of 8\,$h^{-1}$Mpc which measured the normalisation of the mass
power spectrum).  For clusters with mass $\gsim10^{14.5}$, the cluster
abundance is expected to rise by a factor of 6$\times$ and 20$\times$
between $\sigma_{8}$=0.7 and 0.8 and $\sigma_{8}=0.7$ and 0.9
respectively.  Thus, once complete, the DXS has the opportunity to use
galaxy cluster abundances as a precision tool for cosmology and we look
forward to undertaking this task in the future.

The discovery of a supercluster in the first DXS field highlights the
importance of the combination of depth and area in surveys of the
$z=0.5$--2 Universe.  Surveys of one WFCAM field or less are unlikely
to contain a structure as rare as a supercluster or, even if they do,
will only cover part of it.  As such, surveys of this size are still
affected by cosmic variance, and any attempt to measure cosmological
parameters are severely compromised (e.g.  \citealt{Retzlaff98}).  Only
contiguous surveys of several degrees (such as the UKIDSS/DXS,
VISTA/VIDEO and VISTA/VIKING), will have sufficient area and depth
coverage to identify large structures in statistically significant
numbers whilst reliably accounting for the effects of cosmic variance
(see \citealt{Borgani06} for a review).

Indeed, the implications of how cosmic variance affects the analysis of
cluster surveys of relatively small volumes and/or many non-contiguous
areas are subtle but in an era of ``precision cosmology'' must be
considered (e.g.\, \citealt{Schuecker01}).

\section*{acknowledgements}   
We gratefully acknowledge the anonymous referee for their suggestions
which improved the content and clarity of this paper.  We would also
like to thank Vince Eke, Gus Evrard, Adrian Jenkins and Inger
J{\o}rgensen for useful discussions.  AMS and CJS acknowledge PPARC
Fellowships, ACE and IRS acknowledge support from the Royal Society.
We gratefully acknowledge the UKIDSS DXS.  The United Kingdom Infrared
Telescope which is operated by the JAC on behalf of PPARC.  The GMOS
observations were taken as part of programme GN-2006A-Q-18 and are
based on observations obtained at the Gemini Observatory, which is
operated by the Association of Universities for Research in Astronomy,
Inc., under a cooperative agreement with the NSF on behalf of the
Gemini partnership: the NSF (United States), PPARC (United Kingdom),
the NRC (Canada), CONICYT (Chile), the ARC (Australia), CNPq (Brazil)
and CONICET (Argentina).  This paper is also partially based on data
collected at Subaru Telescope which is operated by NAO of Japan as well
as observations made with the Spitzer Space Telescope, which is
operated by the Jet Propulsion Laboratory, California Institute of
Technology under a contract with NASA.

\bibliographystyle{apj} 
\bibliography{/Users/ams/Projects/ref}

\newpage

\begin{table*}
\begin{center}
{\scriptsize
{\centerline{\sc Table 2: Member Galaxies}}
\begin{tabular}{lccccccccccccc}
\hline
\noalign{\smallskip}
ID              & R.A.\           & Dec.\          & $I$       & $J$       & $K$    & 3.6$\mu$m & 4.5$\mu$m & 5.8$\mu$m & 24$\mu$m & $z$ & Sp Type \\
                &  $^{h}$ $'$ $''$   & $^{\circ}$ $'$ $''$ \\
\hline\hline
&&&& \multispan2{$z=0.90$ Cluster Galaxies} \\
\hline\hline
DXS1-11$^{[1]}$ & 16\,08\,26.474 & +54\,35\,33.79 & 22.18[01] & 21.03[03] & 20.19[03] & 19.86 & 20.19 &  ...    &   ...   & 0.88853[$-$25,+24] & k     \\
DXS1-12         & 16\,08\,27.386 & +54\,35\,40.38 & 22.25[01] & 20.83[09] & 19.83[01] & 19.18 & 19.68 &  ...    &   ...   & 0.88590[$-$17,+13] & k     \\
DXS1-13         & 16\,08\,27.480 & +54\,35\,49.09 & 22.40[01] & 21.29[04] & 20.22[02] & 19.92 & 20.38 &  ...    &   ...   & 0.89252[$-$13,+13] & k     \\
DXS1-19         & 16\,08\,34.152 & +54\,36\,17.75 & 24.09[03] & 23.23[25] & 22.19[10] &   ... &   ... &  ...    &   ...   & 0.88294[$-$39,+39] & e(a)  \\
DXS1-22         & 16\,08\,27.818 & +54\,38\,00.82 & 21.82[01] & 20.61[02] & 19.76[01] & 19.11 & 19.44 & 19.68   & 17.07   & 0.87894[$-$32,+95] & k     \\
DXS1-25         & 16\,08\,36.910 & +54\,37\,41.77 & 21.16[01] & 20.16[02] & 19.26[01] & 18.82 & 19.34 & 19.64   &   ...   & 0.91071[$-$14,+15] & k     \\
\hline          
DXS2-0          & 16\,08\,05.302 & +54\,45\,29.16 & 22.55[01] & 21.43[05] & 20.11[02] & 19.49 & 19.78 &  ...    & 16.75   & 0.87830[$-$18,+21] & k     \\
DXS2-6          & 16\,08\,21.552 & +54\,44\,53.30 & 20.97[01] & 20.51[02] & 19.82[01] & 19.44 & 19.83 & 19.66   &   ...   & 0.89761[$-$16,+15] & k+a   \\
DXS2-7          & 16\,08\,24.792 & +54\,44\,25.48 & 22.00[01] & 20.94[03] & 19.76[01] & 18.94 & 19.39 &  ...    &   ...   & 0.90421[$-$18,+16] & k+a   \\
DXS2-8          & 16\,08\,25.368 & +54\,44\,35.71 & 22.50[01] & 21.52[05] & 20.39[02] & 20.05 & 20.46 &  ...    &   ...   & 0.90525[$-$37,+33] & e(a)  \\
DXS2-10         & 16\,08\,25.632 & +54\,45\,09.68 & 22.15[01] & 20.76[03] & 20.18[02] & 19.12 & 19.44 & 19.21   & 16.83   & 0.90602[$-$05,+06] & e(c)  \\
DXS2-11         & 16\,08\,29.520 & +54\,44\,41.46 & 22.41[01] & 20.91[03] & 20.15[02] & 19.55 & 19.97 &  ...    &   ...   & 0.90036[$-$45,+35] & k+a   \\
DXS2-12         & 16\,08\,28.992 & +54\,45\,02.95 & 21.60[01] & 20.24[02] & 19.61[10] & 19.23 & 19.72 &  ...    &   ...   & 0.89816[$-$09,+09] & k+a   \\
DXS2-13         & 16\,08\,28.992 & +54\,45\,19.36 & 22.65[02] & 21.16[04] & 20.88[03] & 20.62 & 21.25 &  ...    &   ...   & 0.89878[$-$27,+24] & k     \\
DXS2-15         & 16\,08\,30.190 & +54\,46\,00.41 & 22.71[01] & 21.53[05] & 20.70[03] & 20.39 & 20.78 &  ...    &   ...   & 0.89931[$-$19,+18] & e(c)  \\
\hline          
DXS3-1          & 16\,09\,02.614 & +54\,59\,15.28 & 21.83[01] & 21.06[03] & 20.25[02] & 19.80 & 20.24 &     ... &   ...   & 0.90651[$-$17,+18] & e(c)  \\
DXS3-3          & 16\,09\,02.592 & +54\,58\,39.43 & 22.39[01] & 21.83[07] & 20.81[03] & 20.40 & 21.01 &     ... &   ...   & 0.89953[$-$27,+23] & e(c)  \\
DXS3-7          & 16\,09\,01.344 & +54\,58\,04.15 & 21.84[01] & 20.93[03] & 20.03[01] & 19.64 & 20.00 &     ... &   ...   & 0.90646[$-$20,+19] & e(a)  \\
DXS3-10         & 16\,09\,04.392 & +54\,57\,32.69 & 21.89[01] & 20.76[03] & 19.84[01] & 19.52 & 19.88 &     ... &   ...   & 0.89543[$-$09,+09] & e(a)  \\       
DXS3-11         & 16\,09\,03.288 & +54\,57\,22.03 & 21.74[01] & 20.73[03] & 19.85[01] & 19.34 & 19.60 &     ... &   ...   & 0.90075[$-$13,+13] & e(c)  \\
DXS3-12         & 16\,09\,01.392 & +54\,57\,15.44 & 22.12[01] & 20.99[03] & 19.87[01] & 19.39 & 19.73 &     ... &   ...   & 0.90739[$-$25,+25] & e(a)  \\
DXS3-13         & 16\,08\,59.974 & +54\,57\,11.30 & 21.54[01] & 20.15[02] & 19.06[01] & 18.39 & 18.84 & 19.13   &   ...   & 0.90081[$-$46,+35] & e(c)  \\
DXS3-14         & 16\,08\,56.280 & +54\,57\,09.40 & 22.27[01] & 21.22[04] & 20.20[02] & 19.88 & 20.14 &     ... &   ...   & 0.90097[$-$34,+31] & k+a   \\       
DXS3-15         & 16\,09\,01.224 & +54\,56\,45.24 & 23.01[02] & 21.95[08] & 20.88[03] & 20.61 & 21.09 &     ... &   ...   & 0.90329[$-$45,+41] & k     \\
DXS3-16         & 16\,08\,59.594 & +54\,56\,34.01 & 21.53[01] & 20.18[02] & 19.18[01] & 18.75 & 19.18 & 19.53   &   ...   & 0.90628[$-$10,+11] & k+a   \\
DXS3-17         & 16\,09\,00.168 & +54\,56\,25.59 & 21.61[01] & 20.27[02] & 19.33[01] & 18.90 & 19.28 & 19.30   &   ...   & 0.90896[$-$19,+21] & k     \\
DXS3-18         & 16\,08\,57.386 & +54\,56\,22.30 & 23.09[02] & 22.20[10] & 20.84[03] & 20.37 & 20.85 &     ... &   ...   & 0.90936[$-$49,+55] & e(c)  \\
DXS3-19         & 16\,09\,01.728 & +54\,56\,06.87 & 23.08[02] & 22.29[10] & 20.71[03] & 20.28 & 20.49 &     ... &   ...   & 0.90328[$-$45,+46] & e(a)  \\
DXS3-20         & 16\,09\,03.146 & +54\,55\,53.91 & 22.35[01] & 21.21[04] & 20.17[02] & 19.67 & 19.96 &     ... &   ...   & 0.90581[$-$50,+49] & e(c)  \\
DXS3-22         & 16\,08\,53.738 & +54\,55\,45.01 & 23.21[02] & 22.42[12] & 20.99[03] & 20.96 & 21.52 &     ... &   ...   & 0.90524[$-$12,+12] & e(a)  \\
DXS3-23         & 16\,09\,02.112 & +54\,55\,18.05 & 22.27[01] & 21.05[03] & 19.99[01] & 19.32 & 19.74 &     ... &   ...   & 0.90047[$-$34,+35] & e(c)  \\
DXS3-24         & 16\,08\,56.042 & +54\,55\,20.25 & 22.67[01] & 21.49[05] & 20.29[02] & 20.01 & 20.32 &     ... &   ...   & 0.89286[$-$40,+37] & e(a)  \\
\hline          
DXS4a-3         & 16\,13\,12.792 & +54\,47\,16.30 & 21.90[01] & 20.62[02] & 19.75[01] & 19.14 & 19.45 & 19.52   & 17.53   & 0.88743[$-$48,+39] & k     \\
DXS4a-7         & 16\,13\,05.424 & +54\,46\,46.35 & 22.60[01] & 21.21[04] & 20.39[02] & 20.16 & 20.76 &    ...  &   ...   & 0.88082[$-$40,+34] & k     \\
DXS4a-16        & 16\,12\,55.678 & +54\,45\,51.26 & 22.35[01] & 20.85[03] & 20.12[02] & 19.97 & 20.26 &    ...  &   ...   & 0.88487[$-$24,+21] & k+a   \\
DXS4a-18        & 16\,12\,54.482 & +54\,45\,11.52 & 23.24[02] & 21.96[08] & 21.19[04] &   ... &   ... &    ...  &   ...   & 0.87986[$-$71,+55] & k     \\
DXS4a-21        & 16\,12\,50.736 & +54\,44\,57.00 & 22.08[01] & 20.78[03] & 20.07[01] & 19.57 & 19.85 & 19.85   & 17.33   & 0.90438[$-$21,+20] & e(c)  \\
DXS4a-22        & 16\,12\,42.554 & +54\,46\,23.88 & 22.83[02] & 21.27[04] & 20.48[02] & 20.17 & 20.61 &    ...  &   ...   & 0.88396[$-$50,+62] & k+a   \\
\hline          
DXS5-0          & 16\,10\,32.256 & +54\,59\,12.77 & 22.48[03] & 21.13[03] & 19.77[01] & 19.06 & 19.47 &  ...    &   ...   & 0.90927[$-$42,+25] & k     \\
DXS5-3          & 16\,10\,30.576 & +55\,00\,44.96 & 22.01[01] & 20.75[02] & 20.01[01] & 19.74 & 20.21 &  ...    &   ...   & 0.90944[$-$28,+28] & k+a   \\
DXS5-7          & 16\,10\,36.430 & +55\,00\,27.53 & 21.82[01] & 20.82[02] & 20.14[01] & 19.85 & 20.29 &  ...    &   ...   & 0.90038[$-$85,+59] & k+a   \\
DXS5-8          & 16\,10\,31.464 & +55\,01\,55.38 & 23.17[01] & 22.65[15] & 20.66[02] & 20.21 & 20.47 & 19.76   &   ...   & 0.90946[$-$25,+25] & e(c)  \\
DXS5-10         & 16\,10\,40.130 & +55\,00\,58.61 & 22.17[01] & 21.10[03] & 20.17[01] & 19.88 & 20.16 &  ...    &   ...   & 0.89330[$-$42,+42] & e(c)  \\
DXS5-13         & 16\,10\,42.122 & +55\,01\,20.06 & 22.05[01] & 20.89[03] & 20.04[01] & 19.64 & 20.02 & 20.22   &   ...   & 0.89137[$-$10,+08] & e(c)  \\
DXS5-14         & 16\,10\,42.792 & +55\,01\,26.68 & 22.12[02] & 20.47[02] & 19.45[00] & 19.03 & 19.41 & 19.84   &   ...   & 0.89043[$-$18,+23] & e(c)  \\
DXS5-16         & 16\,10\,43.968 & +55\,01\,36.12 & 23.34[02] & 22.09[08] & 21.09[03] & 20.74 & 21.04 &  ...    &   ...   & 0.88947[$-$71,+71] & k+a   \\
DXS5-17         & 16\,10\,42.840 & +55\,02\,07.11 & 22.19[01] & 21.00[03] & 20.09[01] & 19.78 & 20.14 &  ...    &   ...   & 0.89275[$-$56,+56] & e(c)  \\
DXS5-20         & 16\,10\,51.624 & +55\,00\,48.96 & 22.68[01] & 21.33[04] & 20.64[02] & 20.54 & 20.91 &  ...    &   ...   & 0.90677[$-$10,+10] & k     \\
DXS5-23         & 16\,10\,56.426 & +55\,00\,41.40 & 22.68[01] & 22.21[10] & 20.80[02] & 20.53 & 21.01 &  ...    &   ...   & 0.89174[$-$61,+61] & k+a  \\
DXS5-24         & 16\,10\,56.232 & +55\,01\,05.45 & 22.07[02] & 21.08[03] & 20.03[01] & 19.56 & 19.92 &  ...    &   ...   & 0.89869[$-$17,+13] & k  \\
\hline\hline
&&&& \multispan2{$z=1.09$ Cluster Galaxies} \\
\hline\hline
DXS4b-4         & 16\,13\,07.752 & +54\,47\,55.60 & 21.62[01] & 21.56[06] & 20.96[03] & 20.20 & 20.48 &   ...   &   ...   & 1.09347[$-$45,+53] & k     \\
DXS4b-8         & 16\,13\,03.264 & +54\,46\,41.16 & 22.84[02] & 20.98[03] & 20.14[02] & 19.71 & 19.95 &   ...   &   ...   & 1.08570[$-$26,+27] & k     \\
DXS4b-11$^{[4]}$& 16\,13\,01.680 & +54\,46\,09.98 & 21.87[01] & 20.31[02] & 19.28[01] & 18.53 & 18.74 & 19.04   & 16.20   & 1.09168[$-$07,+09] & e(c)  \\
DXS4b-12        & 16\,13\,01.270 & +54\,46\,02.06 & 23.17[02] & 21.04[03] & 20.27[02] & 19.72 & 19.93 &   ...   &   ...   & 1.08921[$-$52,+38] & e(a)  \\
DXS4b-13        & 16\,12\,58.846 & +54\,46\,20.00 & 23.36[02] & 22.35[11] & 21.62[06] &   ... &  ...  &   ...   &   ...   & 1.09052[$-$55,+72] & e(b)  \\
DXS4b-14        & 16\,12\,57.840 & +54\,46\,05.19 & 22.22[01] & 21.19[04] & 20.32[02] & 19.63 & 19.88 &   ...   &   ...   & 1.09105[$-$27,+26] & k+a   \\
DXS4b-17        & 16\,12\,53.112 & +54\,45\,42.74 & 23.31[02] & 21.23[04] & 20.69[03] & 20.30 & 20.55 &   ...   &   ...   & 1.09789[$-$14,+13] & k+a   \\
DXS4b-19        & 16\,12\,50.782 & +54\,45\,39.14 & 22.12[01] & 20.55[02] & 19.59[01] & 19.00 & 19.25 & 19.69   &   ...   & 1.09690[$-$04,+09] & e(a)  \\
\hline\hline
\label{tab:DXS1-5a}
\end{tabular}
}
\caption{Note: The RA and Dec are in the J2000 co-ordinate system.  The
  value in the [] is the error in the last decimal place and we note
  that 1\,$\mu$Jy corresponds to m$_{AB}$=23.90. {\bf $^[1]$} {\it
    Lenses:} We have searched for any unidentified lines in the spectra
  to search for lensed background galaxies and identify one candidate
  from the cluster and two from the field population.  DXS1-11 has
  unambiguous identification of [O{\sc ii}]3727, Ca\,H\&K and G-band
  for $z=$0.89, but we also identify weak line emission at 8594.5
  (spatial offset $<$0.5$''$) which we tentatively identify as [O{\sc
    ii}]3727 at $z$=1.306. {\bf $^[2]$} {\it Noteworthy Objects:} We
  note that the brightest cluster member in DXS4b at $z=$1.09 has a
  number of high ionisation lines such as [Ne{\sc v}]3346,3426 and
  [Ne{\sc iii}]3343,3868.7.  The strong 24$\mu$m detection suggests
  that this galaxy contains a highly obscured AGN, although it has a
  low infrared luminosity ($\lsim$5$\times$10$^{11}$L$_{\odot}$ from
  the lack of a 70$\mu$m detection).}
\end{center}
\end{table*}

\begin{table*}
\begin{center}
{\scriptsize
{\centerline{\sc Table 3: Non-member Galaxies}}
\begin{tabular}{lcccccccccc}
\hline
\noalign{\smallskip}
ID              & R.A.\           & Dec.\          & $I$       & $J$       & $K$    & 3.6$\mu$m & 4.5$\mu$m & 5.8$\mu$m & 24$\mu$m & $z$ \\
                & $^{h}$ $'$ $''$     & $^{\circ}$ $'$ $''$ \\
\hline\hline    
DXS1-0          & 16\,08\,18.842 & +54\,33\,30.28 & 22.38[01] & 21.64[06] & 21.03[04] &   ... &   ... &   ...   &   ...   & 1.057 \\
DXS1-1          & 16\,08\,20.928 & +54\,33\,25.20 & 21.75[01] & 20.65[02] & 19.95[01] & 19.65 & 20.14 &   ...   &   ...   & 0.652 \\
DXS1-2          & 16\,08\,15.696 & +54\,34\,13.08 & 20.43[00] & 19.89[01] & 19.05[01] & 19.48 & 19.53 &   ...   &   ...   & 0.322 \\
DXS1-3          & 16\,08\,12.576 & +54\,34\,43.86 & 23.45[02] & 22.04[09] & 20.82[03] & 19.91 & 19.98 &   ...   &   ...   &  ...  \\
DXS1-4          & 16\,08\,20.592 & +54\,34\,48.72 & 21.85[01] & 20.81[03] & 19.97[01] & 19.81 & 20.46 &   ...   &   ...   & 0.780 \\
DXS1-5          & 16\,08\,17.498 & +54\,35\,22.80 & 23.51[02] & 21.76[07] & 20.43[02] & 19.81 & 19.77 &   ...   &   ...   & 1.143 \\
DXS1-6          & 16\,08\,15.506 & +54\,35\,49.02 & 22.28[01] & 20.91[03] & 19.94[01] & 19.64 & 19.95 & 20.02   &   ...   & 1.097 \\
DXS1-7          & 16\,08\,23.470 & +54\,35\,05.24 & 21.68[01] & 20.99[03] & 20.26[02] & 20.03 & 20.55 &   ...   &   ...   & 0.663 \\
DXS1-8          & 16\,08\,26.426 & +54\,34\,52.50 & 22.19[01] & 21.51[05] & 20.53[02] & 20.25 & 20.64 &   ...   &   ...   &  ...  \\
DXS1-9          & 16\,08\,16.464 & +54\,36\,14.00 & 20.81[01] & 19.91[01] & 19.12[01] & 18.31 & 18.60 & 18.56   & 16.19   &  ...  \\
DXS1-10         & 16\,08\,20.544 & +54\,35\,54.02 & 22.13[01] & 21.83[07] & 20.93[03] &   ... &   ... &   ...   &   ...   &  ...  \\
DXS1-14         & 16\,08\,28.512 & +54\,35\,53.66 & 22.43[01] & 22.11[09] & 21.25[04] & 21.04 & 21.35 &   ...   &   ...   & 1.030 \\
DXS1-15         & 16\,08\,27.554 & +54\,36\,09.94 & 21.50[01] & 20.62[02] & 19.63[01] & 19.40 & 19.85 &   ...   &   ...   & 0.799 \\
DXS1-16         & 16\,08\,28.846 & +54\,36\,08.38 & 23.22[02] & 23.08[22] & 21.26[04] &   ... &   ... &   ...   &   ...   &  ...  \\
DXS1-17         & 16\,08\,29.880 & +54\,36\,11.02 & 22.93[02] & 21.68[06] & 20.33[02] & 19.68 & 19.58 &   ...   &   ...   &  ...  \\
DXS1-18         & 16\,08\,34.006 & +54\,36\,05.54 & 22.67[01] & 22.51[10] & 21.28[04] & 20.98 & 21.30 &   ...   &   ...   & 0.606 \\
DXS1-20         & 16\,08\,19.200 & +54\,38\,15.11 & 22.64[01] & 21.52[05] & 19.85[01] & 18.85 & 18.49 & 17.98   & 15.65   & 1.225 \\
DXS1-21         & 16\,08\,20.566 & +54\,38\,18.06 & 22.14[01] & 20.62[02] & 19.72[01] & 19.20 & 19.39 &   ...   &   ...   & 1.135 \\
DXS1-23         & 16\,08\,31.850 & +54\,37\,51.16 & 21.72[01] & 20.61[02] & 19.60[10] & 19.29 & 19.68 &   ...   &   ...   &  ...  \\
DXS1-24         & 16\,08\,43.104 & +54\,36\,41.22 & 22.43[01] & 21.63[06] & 20.98[03] & 20.65 & 20.98 &   ...   &   ...   & 1.021 \\
\hline          
DXS2-1          & 16\,08\,10.752 & +54\,44\,21.92 & 23.51[03] & 21.36[05] & 20.93[03] & 20.37 & 20.58 &   ...   &   ...   & 1.156 \\
DXS2-2          & 16\,08\,19.680 & +54\,42\,41.01 & 22.20[01] & 20.77[03] & 19.83[01] & 19.04 & 19.33 & 19.55   & 16.68   & 1.020 \\
DXS2-3          & 16\,08\,17.930 & +54\,43\,48.53 & 22.04[01] & 20.97[03] & 20.10[02] & 19.73 & 20.10 &   ...   &   ...   & 0.693 \\
DXS2-4          & 16\,08\,19.152 & +54\,43\,57.72 & 23.23[02] & 22.85[19] & 21.54[06] & 21.21 & 21.13 &   ...   &   ...   &   ... \\
DXS2-5          & 16\,08\,21.600 & +54\,44\,14.93 & 22.17[01] & 21.35[05] & 20.31[02] & 19.99 & 20.49 &   ...   &   ...   & 0.670 \\
DXS2-9$^{[3]}$  & 16\,08\,24.096 & +54\,45\,12.52 & 21.57[01] & 20.79[03] & 20.58[02] & 20.43 & 21.14 &   ...   &   ...   & 0.674 \\
DXS2-14         & 16\,08\,33.936 & +54\,44\,28.94 & 22.84[02] & 21.62[06] & 20.73[03] & 20.17 & 20.54 &  ...    &   ...   & 0.763 \\
DXS2-16         & 16\,08\,32.112 & +54\,45\,49.02 & 23.79[03] & 21.92[08] & 21.37[05] & 20.83 & 20.88 &   ...   &   ...   & 1.329 \\
DXS2-17         & 16\,08\,37.680 & +54\,45\,05.46 & 22.17[01] & 21.47[05] & 21.04[04] & 20.98 & 21.14 &   ...   &   ...   & 0.734 \\
DXS2-18         & 16\,08\,33.120 & +54\,46\,34.10 & 22.82[02] & 22.10[10] & 21.16[04] &   ... &   ... &   ...   &   ...   & 0.763 \\
DXS2-19         & 16\,08\,40.536 & +54\,45\,13.79 & 23.82[03] & 21.40[05] & 20.99[03] & 20.56 & 20.86 &   ...   &   ...   &   ... \\
DXS2-20         & 16\,08\,44.088 & +54\,44\,44.46 & 22.00[01] & 20.90[03] & 20.33[02] & 20.14 & 20.75 &   ...   &   ...   & 0.800 \\
DXS2-21         & 16\,08\,47.498 & +54\,44\,13.13 & 21.93[01] & 20.48[02] & 19.88[01] & 19.62 & 20.09 &   ...   &   ...   & 0.775 \\
DXS2-22         & 16\,08\,38.448 & +54\,46\,49.88 & 22.83[02] & 21.80[07] & 20.59[02] & 20.00 & 20.43 &   ...   &   ...   & 0.812 \\
DXS2-23         & 16\,08\,41.064 & +54\,46\,26.15 & 22.80[02] & 21.75[07] & 21.12[04] & 20.46 & 20.44 &   ...   &   ...   & 0.641 \\
DXS2-24         & 16\,08\,40.346 & +54\,46\,57.14 & 21.93[01] & 20.83[03] & 19.91[01] & 19.48 & 19.99 &   ...   &   ...   & 0.802 \\
\hline          
DXS3-0          & 16\,09\,08.738 & +54\,59\,16.02 & 22.92[02] & 22.00[08] & 21.17[04] & 20.58 & 20.72 &   ...   &   ...   & 1.3348 \\ 
DXS3-2          & 16\,09\,16.102 & +54\,58\,45.26 & 22.11[01] & 21.68[06] & 20.73[03] & 20.85 & 21.58 &   ...   &   ...   & 0.9510 \\ 
DXS3-4          & 16\,09\,00.648 & +54\,58\,31.30 & 22.67[01] & 21.67[06] & 20.59[03] & 20.54 & 20.92 &   ...   &   ...   & 0.6841 \\ 
DXS3-5          & 16\,09\,07.658 & +54\,58\,13.54 & 22.73[01] & 21.90[08] & 20.53[02] & 20.05 & 20.23 &   ...   &   ...   & 0.9342 \\ 
DXS3-6          & 16\,08\,53.016 & +54\,58\,26.08 & 21.90[01] & 21.64[06] & 20.78[03] & 21.17 & 21.23 &   ...   &   ...   & 1.3155 \\ 
DXS3-8          & 16\,09\,00.816 & +54\,57\,57.13 & 21.26[01] & 20.26[02] & 19.50[01] & 19.66 & 19.92 &   ...   &   ...   &  ...   \\ 
DXS3-9          & 16\,08\,59.062 & +54\,57\,48.25 & 23.56[03] & 22.05[09] & 21.08[04] & 20.61 & 21.10 &   ...   &   ...   & 1.2889 \\ 
DXS3-21         & 16\,08\,51.888 & +54\,55\,59.84 & 22.49[01] & 22.19[10] & 20.68[03] & 20.43 & 20.33 & 19.80   &   ...   & 1.3910 \\   
DXS3-25         & 16\,08\,53.686 & +54\,55\,17.18 & 22.66[01] & 21.76[07] & 20.53[02] & 20.01 & 19.81 & 19.77   & 17.19   &   ...  \\  
DXS3-26         & 16\,08\,51.504 & +54\,54\,58.25 & 22.24[01] & 21.97[08] & 21.05[04] &   ... &   ... &   ...   &   ...   & 0.2999 \\ 
DXS3-27         & 16\,09\,00.768 & +54\,54\,35.24 & 23.16[02] & 21.66[06] & 20.06[01] & 19.17 & 19.24 & 19.46   & 17.47   &   ...  \\
DXS3-28$^{[4]}$ & 16\,08\,52.704 & +54\,54\,32.76 & 21.82[01] & 21.16[04] & 20.20[02] & 20.33 & 20.72 &   ...   &   ...   & 0.6675 \\
\hline          
DXS4-0          & 16\,13\,15.600 & +54\,47\,46.47 & 22.90[02] & 21.54[05] & 21.15[04] &   ... &  ...  &   ...   &   ...   & 0.758 \\
DXS4-1          & 16\,13\,16.128 & +54\,47\,29.00 & 22.84[02] & 20.64[02] & 19.89[01] & 19.33 & 19.43 &   ...   &   ...   & 1.332 \\
DXS4-2          & 16\,13\,13.968 & +54\,47\,29.98 & 22.29[01] & 21.92[08] & 22.02[09] &   ... &  ...  &   ...   &   ...   & 0.644 \\
DXS4-5          & 16\,13\,10.200 & +54\,46\,59.09 & 21.56[01] & 20.89[03] & 20.25[02] & 20.42 & 20.41 &   ...   &   ...   & 0.493 \\
DXS4-6          & 16\,13\,00.792 & +54\,48\,01.26 & 23.46[02] & 21.39[05] & 21.11[04] & 20.40 & 20.67 &   ...   &   ...   & 1.309 \\
DXS4-9          & 16\,13\,04.754 & +54\,45\,55.69 & 23.30[02] & 21.66[06] & 20.58[02] & 19.90 & 19.97 &   ...   &   ...   & ...   \\
DXS4-10         & 16\,13\,00.792 & +54\,46\,33.54 & 23.51[02] & 21.06[03] & 20.17[02] & 19.31 & 19.39 &   ...   &   ...   & ...   \\
DXS4-15         & 16\,12\,54.240 & +54\,46\,26.97 & 21.06[01] & 20.35[02] & 19.78[01] & 20.16 & 20.37 &   ...   &   ...   & 0.453 \\
DXS4-20         & 16\,12\,50.618 & +54\,45\,22.54 & 21.45[01] & 20.57[02] & 19.63[01] & 19.66 & 20.15 &   ...   &   ...   & 0.600 \\
DXS4-23         & 16\,12\,46.586 & +54\,45\,18.29 & 23.67[02] & 21.50[05] & 20.42[02] & 19.63 & 19.51 & 19.56   &   ...   & ...   \\
DXS4-24         & 16\,12\,46.802 & +54\,44\,56.65 & 23.41[02] & 21.42[05] & 20.76[03] & 20.37 & 20.37 &   ...   &   ...   & 1.344 \\
\hline                                                                                                                             
DXS5-1          & 16\,10\,26.880 & +55\,00\,58.57 & 22.78[01] & 21.14[03] & 20.07[01] & 19.50 & 19.75 &  ...    &   ...   &  ...  \\
DXS5-2          & 16\,10\,34.776 & +54\,59\,22.17 & 22.12[01] & 21.28[04] & 20.54[02] & 20.86 & 21.50 &  ...    &   ...   & 0.634 \\
DXS5-4          & 16\,10\,36.098 & +54\,59\,35.98 & 22.94[06] & 22.10[09] & 21.12[03] &   ... &   ... &  ...    &   ...   & 0.807 \\
DXS5-5          & 16\,10\,40.224 & +54\,58\,48.64 & 21.65[01] & 20.76[02] & 19.76[01] & 19.60 & 20.17 &  ...    &   ...   & 0.807 \\
DXS5-6          & 16\,10\,28.730 & +55\,02\,02.54 & 22.85[01] & 21.31[04] & 20.04[01] & 19.62 & 19.78 &  ...    &   ...   &  ...  \\
DXS5-9          & 16\,10\,32.978 & +55\,02\,18.50 & 23.46[01] & 21.46[05] & 21.17[04] &   ... &   ... &  ...    &   ...   &  ...  \\
DXS5-11         & 16\,10\,45.576 & +54\,59\,54.78 & 23.67[01] & 21.76[06] & 20.79[02] & 20.22 & 20.15 &  ...    &   ...   &  ...  \\
DXS5-12         & 16\,10\,44.616 & +55\,00\,22.25 & 23.38[01] & 22.34[11] & 21.22[04] & 21.26 & 21.51 &  ...    &   ...   &  ...  \\
DXS5-15         & 16\,10\,43.488 & +55\,01\,29.56 & 22.67[01] & 21.27[04] & 20.24[01] & 19.32 & 19.62 & 19.92   & 17.13   &  ...  \\
DXS5-18         & 16\,10\,48.432 & +55\,00\,57.10 & 21.72[01] & 20.63[02] & 19.58[00] & 19.36 & 19.88 & 19.94   &   ...   & 0.713 \\
DXS5-21         & 16\,10\,47.878 & +55\,02\,14.35 & 23.68[01] & 21.50[05] & 21.04[03] & 20.28 & 20.11 & 19.85   &   ...   & 1.33  \\
DXS5-22         & 16\,10\,51.240 & +55\,01\,44.11 & 23.81[01] & 22.91[19] & 22.46[13] &   ... &   ... &  ...    &   ...   & 1.13  \\
DXS5-25         & 16\,10\,54.744 & +55\,02\,09.42 & 21.85[01] & 21.02[03] & 20.07[01] & 19.88 & 20.24 &  ...    &   ...   & 0.806 \\
DXS5-26         & 16\,10\,57.770 & +55\,01\,47.28 & 23.47[01] & 22.36[11] & 20.70[02] & 20.06 & 20.01 & 20.00   &   ...   &  ...  \\
DXS5-27         & 16\,10\,52.512 & +55\,03\,20.84 & 23.60[02] & 22.35[11] & 20.67[02] & 20.29 & 20.61 &  ...    &   ...   &  ...  \\
DXS5-28         & 16\,10\,59.378 & +55\,02\,28.18 & 22.86[01] & 22.91[19] & 21.20[04] & 21.61 & 21.72 &  ...    &   ...   &  ...  \\
\hline\hline
\label{tab:DXS1-5b}
\end{tabular}
}
\vspace{-0.5cm}
\caption{Note: {\it Lenses:} $^{[3]}$ DXS2-9 is a $z=$0.67 galaxy
  through identification of Ca\,H\&K, H$\beta$4861 and [O{\sc
    iii}]4959,5007.  However we also identify [O{\sc ii}]3727 and
  Ca\,H\&K at 8109, 8555 and 8632\AA\ respectively (spatial offset
  $<$0.5$''$) giving a redshift of $z=$1.175.  $^{[4]}$ DXS3-28 is
  identified as a $z=$0.67 galaxy from [O{\sc ii}]3727 emission and
  Ca\,H\&K absorption.  However, spatially offset by $\sim$0.75$''$ to
  the south we also identify line emission at 6748,8801,8979\&9067\AA\
  which corresponds to redshifted [O{\sc ii}]3727, H$\beta$ and [O{\sc
    iii}]4959,5007 for $z=0.81$}.
\end{center}
\end{table*}

\end{document}